\documentclass[10pt]{iopart}
\usepackage{epsfig,graphics,color}
\usepackage{graphicx}
\newcommand \etc {{\it etc.}}
\newcommand \tie {{\it i.e.}}

\newcommand \ra  {\rightarrow}

\newcommand \B {\beta}

\newcommand \lc {\langle}
\newcommand \rc {\rangle}
\newcommand \prt {\partial}

\newcommand \sg {\sigma}

\newcommand \nt {\noindent}

\newcommand \bvec{\left( \begin{array}{c} }
\newcommand \evec{\end{array} \right)}

\newcommand \bea{\begin{eqnarray} }
\newcommand \eea{\end{eqnarray} }

\newcommand {\be} {\begin{equation}}
\newcommand {\ee} {\end{equation}}

\newcommand{\sq}{\sigma^2}

\begin{document}

\title[Correlations of Conserved Charges]{Deciphering Deconfinement in Correlations of Conserved Charges}

\author{A. Majumder}
\address{Department of Physics, Duke University, Durham, NC 27708-0305, USA}

\date{ \today}

\begin{abstract}
Diagonal and off-diagonal flavor and conserved charge susceptibilities reveal the prevalent degrees of freedom of heated strongly interacting matter. Results obtained from lattice simulations are compared with various model estimates in an effort to weed down various possible pictures of a quark gluon plasma. We argue that the vanishing of the off-diagonal quark flavor susceptibilities and its derivatives with respect to chemical potential, at temperatures above 1.5 Tc, can only be understood in a picture of a gas or liquid composed of quasi-particles which carry the same quantum numbers as quarks and antiquarks. A potential new observable, blind to neutral and non-strange particles, is introduced and related via isospin symmetry to the ratio of susceptibilities of baryonic strangeness to strangeness generated in the excited matter created at RHIC.
\end{abstract}


The goal of of the heavy-ion program at the Relativistic Heavy Ion Collider 
(RHIC)  is the creation and study of heated strongly interacting matter at 
nearly vanishing baryon density \cite{Harris:1996zx,Gyulassy:2004zy}. 
The experimental results of the four RHIC detector collaborations, have  
set a lower bound of about 5 GeV/fm$^3$ on the energy density at a time 
$\tau = 1$ fm/c in central Au+Au collisions \cite{RHIC_Whitepapers}.
Such an energy density should place the produced 
matter well into region of the QCD phase diagram that cannot be described
as a hadronic resonance gas. Matter in this domain of the QCD phase diagram 
had been expected to be a
colored plasma composed of quasiparticle excitations with the quantum 
numbers of quarks and gluons \cite{Harris:1996zx,Rischke:2003mt}.
The produced matter, however, 
clearly exhibits collective behavior as evidenced by its radial and elliptic 
flow \cite{Ackermann:2000tr}. 
This result suggests that there must exist a strong 
interaction between the constituents of the medium. 

Lattice simulations of QCD at finite temperature and vanishing chemical 
potential have demonstrated the existence of a phase transition at $T=T_c \simeq 170$MeV
signalled by a steep rise in the energy density and the pressure 
as a function of the temperature. The slow rise of both quantities prior to the 
sudden transition has been successfully modelled by a picture of a hadronic 
resonance gas \cite{Karsch:2003zq}. A similar program for the description of 
the excited phase as a weakly interacting plasma of quasiparticles 
\cite{Andersen:1999va} has not met with success in the region 
from $T_c \leq T \leq 3T_c$. 
This indicates that matter in this region may not be a weakly 
coupled plasma where quarks and gluons are deconfined over large distances. 
It is clear that 
a microscopic understanding of the emergent degrees of freedom in this regime 
 is essential in order to understand the matter created in nuclear collisions at RHIC. 
The object of  these proceedings is to outline a set of 
flavor off-diagonal susceptibilities which may be calculated on the lattice, in phenomenological models and 
in certain situations may even be measured in experiments. Such off-diagonal susceptibilities 
are discerning probes of  the thermodynamic degrees of freedom in 
strongly interacting QCD matter, both above and below $T_c$. 

In the lattice formulation of QCD, the fundamental degrees of freedom
are local quark and gluon fields. Under conditions where deconfinement 
has been achieved, the elementary set of conserved charges is given by
the net contents of each quark flavor $u,d,s$.
An alternate basis is provided by the hadronically 
defined conserved charges of baryon number ($B$), electric charge ($Q$) 
and strangeness ($S$). 
The (co-)variances between any two conserved charges ($x,y$) are extensive quantities: 
$
\sq_{xy} = VT \chi_{xy}, 
$
where $\chi_{xy}$ is the intensive diagonal or off-diagonal susceptibility.
These susceptibilities can be measured on the lattice. In heavy-ion experiments, 
the variances and covariances are measured by means of an event-by-event 
analysis of the corresponding conserved quantities, \tie,
\bea
\sq_{xy} = \frac{1}{N_{E}}\sum_{i \in E} X_i Y_i - 
\left(\frac{\sum_{i \in E} X_i}{N_{E}}\right) 
\left(\frac{\sum_{i \in E} Y_i}{N_{E}}\right),
\eea
where $E$ represents the set of events, $N_E$ is the number of events
considered, and $X_i,Y_i$ are the net values of the conserved charge in 
a given event $i$. The volume independent ratios of variances, measured 
event-by-event in heavy-ion collisions, may then be directly compared 
with the lattice estimate for the ratio of susceptibilities. 

In the case that the degrees of freedom or active 
flavors $f$  are eigenstates of  the quantum numbers $x,y$, the above expression may 
be decomposed using $X_i = \sum_f n_i^f x_f $ where $n_i^f$ are the number of 
states of flavor $f$ in event  $i$. For independent flavors, where 
$\lc n^f n^g \rc = \sum_{i \in E} n_i^f  n_i^g =  \lc n^f \rc \lc n^g \rc$, one 
may easily demonstrate that (see Ref.~\cite{Majumder:2006nq} for details)
\bea
\sq_{xy} &=& \sum_f \sq_f  x_f  y_f  =\!\!\!\!\!\!\!\!\mbox{}^{P.S.} \sum_f  \lc n^f \rc x_f y_f. \label{sg_qs2}
\eea
The last equality in the above equation holds solely in the case that Poisson 
statistics is applicable to the independent flavors (\tie, the variance $\sq$ 
is equal to the mean $\lc n \rc$) 
Equation \eref{sg_qs2} clearly demonstrates 
the sensitivity of the off-diagonal susceptibilities to the presence of flavors carrying more than one
quantum number, and may be used to estimate the susceptibilities in various microscopic models. 

\begin{figure}
\resizebox{2.5in}{2.5in}{\includegraphics[0.5in,0.5in][5in,5in]{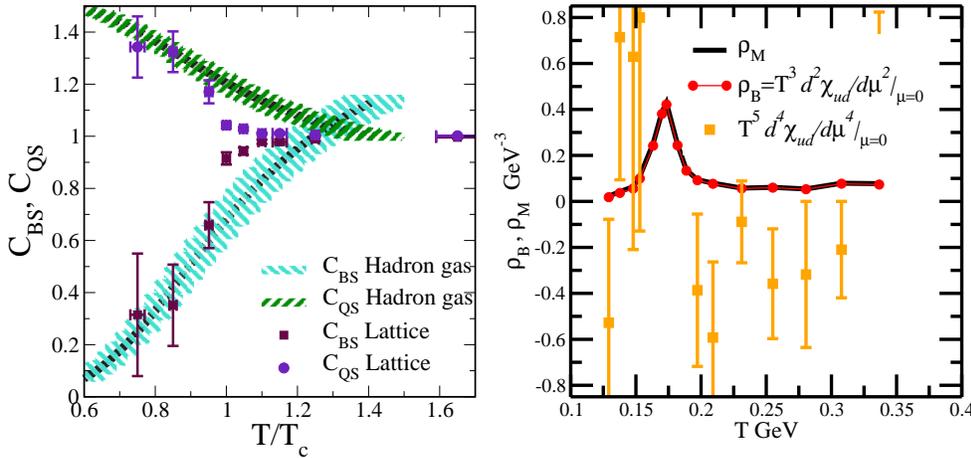}}
\resizebox{2.5in}{2.5in}{\includegraphics[0.5in,0.5in][5in,5in]{c_ud.eps}}
\caption{(Left Panel) A comparison of  $C_{BS}$, $C_{QS}$ calculated in a truncated hadron 
resonance gas at $\mu_B=\mu_S=\mu_Q=0$MeV compared to lattice calculations from 
Ref.~\cite{Gavai:2005yk}. The two hazed bands for $C_{BS}$ and $C_{QS}$ for the hadron gas 
plots reflect the uncertainty in the actual value of the phase transition temperature $T_c = 170 \pm 10$MeV.\\
(Right Panel) A plot of the mesonic and baryonic contributions to $\chi_{ud}(T,\mu=0)$ and 
and a plot of the derivatives $T\prt^2 \chi_{ud} / \prt (\mu \B)^2|_{\mu=0}$ (which 
is equal to the baryonic contribution $\rho_B$) and $T\prt^4 \chi_{ud} / \prt (\mu \B)^4|_{\mu=0}$ 
(which is negative beyond $T_c = 170$MeV and thus is inconsistent with a bound 
state  interpretation).}
\label{fig1b}
\end{figure}

A robust set of ratios of susceptibilities, which has been measured on the lattice 
and demonstrates direct sensitivity to the flavor carrying degrees of freedom
are the two coefficients (see Refs.~\cite{Koch:2005vg,Gavai:2005yk})
 $C_{BS} = -3 \sq_{BS}/\sq_S$  and $C_{QS} = 3 \sq_{QS}/\sq_S$. 
The susceptibilities may be computed phenomenologically, 
assuming the condition of Eqs.~\eref{sg_qs2}.
Plotted in the left panel of Fig.~\ref{fig1b}. are 
the coefficients $C_{BS}$ and $C_{QS}$ as obtained from a hadronic 
resonance gas spectrum, truncated at the mass of the $\Omega^-$ along with 
the results obtained from the lattice. 
One notes, that the hadron resonance gas provides a good description of 
the behavior of the ratio of susceptibilities  up to the point of the phase transition. 
Here the behavior of the truncated spectrum fails to reproduce the sharp rise in $C_{BS}$  
and the corresponding sharp drop in $C_{QS}$. 

The results from the lattice simplify  in the high temperature phase, where 
one notes the following simple relations \cite{Gavai:2002jt,Allton:2005gk}; 

\bea
 \chi_{us} = \chi_{ds} \leq \chi_{ud}  \ll \chi_{s} \leq \chi_{d} = \chi_{u},
\eea

\nt
where the last and the first equality applies in the case of isospin symmetry. 
For simulations at vanishing chemical potentials $\lc B \rc = \lc S \rc = \lc Q \rc = 0$, as a result, we 
obtain, 
\bea
C_{BS} = -3 \frac{\lc B S \rc}{\lc S^2 \rc} 
= \frac{\lc( u + d + s ) s \rc  }{\lc s^2 \rc} \approx  
\frac{\chi_{us} + \chi_{ds} + \chi_s }{\chi_s}  \approx 1.
\label{constraint}
\eea
Alternatively, one may state that $\chi_{us} \simeq \chi_{ds} \ra 0$.
Models of the deconfined phase must obey such constraints. The simplest 
model of deconfined matter is that of non-interacting quark, anti-quark and 
gluon quasi-particles. In 
such a situation, off-diagonal susceptibilities are identically zero as the 
degrees of freedom may carry only a single flavor. 

Recently, an alternate picture of  the deconfined phase has been proposed, where a tower of 
bound states of quarks and gluons are present besides the quasiparticles themselves 
\cite{Shuryak:2004cy}. 
The large cross-sections from resonance scattering imply short mean 
free paths and lend consistency to the macroscopic hydrodynamic picture henceforth 
used to describe the dynamical evolution of the matter produced at RHIC.  
However, in a plasma containing bound states of quarks, in the form of colored 
mesons, diquarks and quark-gluon bound states, the correlation between flavors 
is no longer negligible. In a 
gas consisting solely of quark-antiquark bound states, built from three flavors 
of quarks, the contributing states are $u\bar{d},d\bar{u},u\bar{s},s\bar{u},d\bar{s},s\bar{d}$ 
(The flavor singlets $q\bar{q}$ and gluon bound states will be ignored as they carry no conserved flavor). 
In such a system, the ratio $C_{BS}$ must vanish, because all states have vanishing
baryon number.  Thus meson like states produce a considerable negative contribution to 
the off-diagonal susceptibilities $\chi_{us},\chi_{ud},\chi_{ds}$ and tend to depress the value of 
$C_{BS}$ from unity. 
Quark gluon bound states contribute similarly as quark-antiquark quasiparticles. 
Inclusion of all such states at a temperature of 
$T=1.5T_c$ yields a value $C_{BS} = 0.62$ (or equivalently, negative 
off-diagonal susceptibilities) quite different from the values   
found on the lattice \cite{Koch:2005vg}. 

These results have motivated the inclusion of a variety of baryonic states into the 
model outlined above \cite{Liao:2005pa}. Baryonic states provide additive 
contributions to the off-diagonal susceptibilities and as a result also to $C_{BS}$. 
The inclusion of a sufficient number of baryonic 
states may lead to a balance between the mesonic and baryonic contributions and 
one may engineer a $\chi_{us} = \chi_{ds} = 0$ and as a result  a $C_{BS} = 1$ 
in agreement with the lattice. To discern between 
these two possibilities, one turns to higher derivatives of the the off-diagonal 
susceptibilities in terms of the baryon-chemical potential $\mu_B$. 

In the interest of 
simplicity, the flavor group will be restricted to SU(2)$_f$. Lattice results for 
susceptibilities and their derivatives with respect to baryon 
chemical potential using dynamical quarks exist in this case \cite{Allton:2005gk}. 
The model will consist of quark and 
anti-quark quasi-particles $u,d$ and $\bar{u},\bar{d}$, meson like bound states 
$u\bar{u},d\bar{d},u\bar{d},d\bar{u}$, diquark states $uu,dd,ud$ and their antiparticles 
as well as baryons $uuu,uud, udd, ddd$ and the corresponding anti-baryons. 
In this simplified situation, one may derive the following simple relations 
between the off-diagonal susceptibility and the meson and baryon densities \cite{Majumder:2006nq} ($\rho^0_x = n^0_x/V$, where 
$n^0_x$ denotes the population of a given flavor at $\mu=0$ and not the difference between the flavor 
anti-flavor populations)
\bea
\!\!\!\!\!\!\!\!\!\!\!\!\!\left.T\chi_{ud}(T,\mu)\right|_{\mu=0} \,\,&\simeq&  -2 \rho^0_{u \bar{d}}(T) 
+ \Large\{ 2 \rho^0_{u d}(T) + 4 \rho^0_{uud}(T) 
+ 4 \rho^0_{udd}(T) \Large\} ,  \label{sus_mu} \\
\!\!\!\!\!\!\!\!\!\!\!\!\!T \left[ \frac{\prt^2 \chi_{ud}}{ \prt (\mu \B)^2} \right]_{\mu = 0} 
&=&  2 \rho^0_{u d}(T) + 4 \rho^0_{uud}(T) + 4 \rho^0_{udd}(T) 
= T \left[ \frac{\prt^4 \chi_{ud}}{ \prt (\mu \B)^4} \right]_{\mu = 0} .
\label{chi2_chi4}
\eea

\nt
In this way, one may divide the contributions to the off-diagonal 
susceptibility and its derivatives in terms of mesonic ($\rho_M = 2 \rho^0_{u \bar{d}}(T)$) 
and baryonic ($\rho_B = 2 \rho^0_{u d}(T) + 4 \rho^0_{uud}(T) + 4 \rho^0_{udd}(T)$) 
contributions. 
Using the measured susceptibility and its derivatives, 
these densities are estimated as a function of the temperature.
and plotted in the right panel of Fig.~\ref{fig1b}. 
These densities satisfy Eq.~\eref{sus_mu} 
and the first equality of Eq.~\eref{chi2_chi4}.
The second condition imposed by Eq.~\eref{chi2_chi4} 
has to be satisfied by the fourth derivative of the susceptibility in such a 
picture of bound states. The fourth derivative of the susceptibility, as measured on the lattice,  
has been plotted as the square symbols. Despite large error bars, one 
notes that while the baryon density or the second derivative of the 
susceptibility is consistent with Eq.~\eref{chi2_chi4} below the phase 
transition temperature, it becomes inconsistent with such a condition 
above the phase transition temperature. Above $T_c$, 
$T \prt^4 \chi_{ud} / \prt (\mu \B)^4$ is actually negative; hence, 
no bound state picture is compatible with such results. 
It has already been pointed out in Ref.~\cite{Allton:2005gk} that the 
signs of the various derivatives of the susceptibility are consistent 
with the picture of a weakly interacting quasi-particle gas.

In the experimental  measurement of baryon-strangeness correlations, any detector has to accurately assess the 
baryon number and strangeness in a given rapidity bin in each event. As most detectors 
are blind to stable uncharged particles they cannot measure the neutron and antineutron populations. 
As a result, a measurement of $\sg_{BS}$ may become rather difficult. 
As a recourse, a  new quantum 
number is constructed,
$
M = B + 2 I_3,
$
and fluctuations of $M$ with respect to $S$ are studied. One notes that the assumption of
isospin symmetry reduces the covariance $\sg_{MS}$ to simply $\sg_{BS}$ (see Ref.~\cite{Majumder:2006nq} for 
details).
As a result, in all theoretical models with isospin symmetry $C_{MS} = C_{BS}$. 
In the experimental determination, $M$ has the advantage that
it is vanishing for all particles that do not carry charge or strangeness,  thus 
$M=0$ for neutrons, antineutrons, neutral pions, \etc. There exists a greater 
possibility of contamination of such a signal from charged pions ($\pi^{\pm}$
do not contribute to the measurement of $C_{BS}$). The amount of contamination 
is assessed by comparing measurements of  $C_{BS}$ and $C_{MS}$ in an identical set 
of  HIJING events~ \cite{Wang:1991ht}.

\begin{figure}
{\centerline{\resizebox{2.2in}{2.2in}{\includegraphics[0.5in,0.5in][5.5in,5.5in]{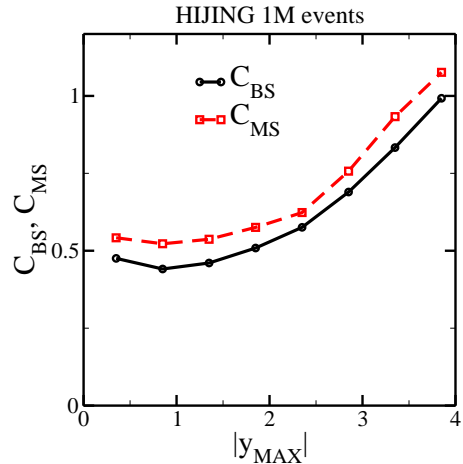}}} }
\caption{A comparison of the two related ratios of variances $C_{BS},C_{MS}$ as a 
function of  the acceptance in rapidity from $-|y_{max}|$ to $|y_{max}|$.}
\label{fig4}
\end{figure}

In Fig.~\ref{fig4} the correlations $C_{BS}$ and $C_{MS}$ are estimated in central 
$Au-Au$ events at $\sqrt{s} = 200$ $A$GeV. 
The acceptance in rapidity ranges from $-|y_{max}|$ to 
$|y_{max}|$, hence a larger $y_{max}$ indicates a larger acceptance. 
The results are presented as a function of $y_{max}$. One notes 
that 
over the range of $y_{max}$ the two correlations $C_{BS}$ and $C_{MS}$ are
rather similar. 
This bodes well for the measurement of $C_{BS}$  in 
RHIC experiments via a measurement of the quantity $C_{MS}$, over a range of 
rapidity intervals.  

In these proceedings, we have demonstrated that the off-diagonal susceptibilities such as 
$\chi_{ud},\chi_{us},\chi_{ds}$ and the ratio $C_{BS}$ are descerning probes of the 
flavor structure  of heated strongly interacting matter. They may be calculated 
on the lattice, in models and even measured in RHIC experiments allowing for a systematic 
exploration of degrees of freedom of deconfined matter. 

Work supported  by the U.S. Department of Energy, grant DE-FG02-05ER41367.

\end{document}